%% file: main.tex
\documentclass[a4paper, amsfonts, amssymb, amsmath, preprint, showkeys, nofootinbib, twoside]{revtex4-1}
\usepackage[english]{babel}
\usepackage[utf8]{inputenc}

\usepackage[colorinlistoftodos, color=green!40, prependcaption]{todonotes}
\input{preamble}
\newcommand{\spr}[1]{{\color{red}\bf[SP:  {#1}]}}
\newcommand{\lb}[1]{{\color{blue}\bf[LB:  {#1}]}}

\usepackage[pdftex, pdftitle={Article}, pdfauthor={Author}]{hyperref} 
\begin{document}
\title{A Louder Gravitational Wave Bang from a Fast-Expanding Universe}

\author{Lucas Brown}
\email{lubrown@ucsc.edu}

\affiliation{Department of Physics, University of California, Santa Cruz (UCSC),
Santa Cruz, CA 95064, USA}
\affiliation{Santa Cruz Institute for Particle Physics (SCIPP),
Santa Cruz, CA 95064, USA}

\author{Aditi Gangadharan}
\email{amg8708@psu.edu}
\affiliation{Department of Physics, University of California, Santa Cruz (UCSC),
Santa Cruz, CA 95064, USA}
\affiliation{Department of Astronomy and Astrophysics, The Pennsylvania State University, State College PA 16801, USA}

\author{Zeynep Su Ko\c c}
\email{su.koc@ug.bilkent.edu.tr}
\affiliation{Department of Physics, University of California, Santa Cruz (UCSC),
Santa Cruz, CA 95064, USA}
\affiliation{Department of Physics, Bilkent University,\\ 06800 Bilkent, Ankara, Turkey}

\author{Stefano Profumo}
    \email[Correspondence email address: ]{profumo@ucsc.edu}
    \affiliation{Department of Physics, University of California, Santa Cruz (UCSC),
Santa Cruz, CA 95064, USA}
\affiliation{Santa Cruz Institute for Particle Physics (SCIPP),
Santa Cruz, CA 95064, USA}


\begin{abstract}
A strong first-order phase transition in a dark sector may produce all or part of the low-frequency gravitational wave signal recently reported by the NANOGrav Collaboration and other pulsar timing arrays. Here we point out, with a simple toy model, that even if the amplitude of the gravitational wave background from the dark phase transition is insufficient to match the NANOGrav signal, a modified expansion rate at early times may considerably enhance the gravitational wave signal. In particular, a faster-than-standard expansion rate, triggered, for instance, by the presence of one or more additional sources of energy density redshifting with higher powers of temperatures than radiation, boosts upper limits on the gravitational wave signal from first-order cosmological phase transitions, enlarging the slate of possible dark sector scenarios matching the NANOGrav signal.

\end{abstract}


\maketitle


\section{Introduction} \label{sec:Introduction}


Gravitational waves (GWs) provide a unique and powerful way to explore the universe's early history, primarily because of their negligible interactions with matter after their formation. High-energy processes that produced GWs in the very early universe should lead to a present-day stochastic background of GW radiation similar in many respects to the cosmic microwave background (CMB) and potentially detectable by current and future GW detectors.
In particular, nanohertz GWs represent a captivating frontier in observational cosmology, offering potential insight into some of the universe's most fundamental processes. These low-frequency GWs, detectable through techniques that rely, for example, on pulsar timing irregularities, have transformed gravitational wave astronomy, opening new windows into the early universe's violent past \cite{ref1,ref4,ref6}. 

Pulsar Timing Arrays (PTAs) stand at the forefront of detecting nanohertz GWs, given their unique ability to measure minute deviations in pulsar timings across the galaxy that might hint at passing gravitational waves \cite{ref1,ref6,ref18}. PTA observations by several collaborations have provided suggestive evidence for a stochastic GW background, sparking significant interest in the question of what could be the source of a nHz-frequency signal. 
\cite{ref4,ref16,ref17,ref18}. Specifically, in 2020, the North American Nanohertz Observatory for Gravitational Waves (NANOGrav) collaboration provided strong evidence for a nHz-frequency common red noise (CRN) process in the residual timing data of pulsars, thought to be a precursor to the detection of a GWB. More recently, in 2023, the inter-pulsar correlation pattern of the CRN signal was shown to follow the Hellings and Down (HD) curve, a telltale sign of a gravitational wave origin \cite{NANOGrav:2023gor}. 

Given this compelling evidence for the detection of a GW background (GWB), there has been a rush to determine which cosmological or astrophysical model sources could reproduce the observed power spectrum of the CRN process \cite{ref4,ref16,ref17,ref18}. One plausible astrophysical source in this frequency range is the in-spiral of numerous supermassive black hole binaries (SMBHBs) throughout the universe, but the local SMBHB density would need to be increased by factors of a few, compared to estimates that account for the amplitude of the observed CRN power spectrum. Alternatively, the signal may have a cosmological origin. Some early universe processes that can generate appreciable GW emission include inflation, first-order phase transitions, cosmic strings, or the formation of primordial black holes.

Here, we focus on the specific possibility that the NANOGrav signal be associated with a first-order phase transition (FOPT) in a ``dark sector''. FOPTs are, generically, pivotal phenomena in cosmology because they can significantly affect the universe's thermal and structural dynamics \cite{Breitbach2018,ref5,ref7}. As the universe cooled, processes akin to the bubbling of boiling water occurred, where bubbles of a new phase nucleated and expanded within the old phase, potentially generating gravitational waves \cite{ref5,ref7}. These transitions are characterized by a latent heat release and the creation of bubble walls moving at relativistic speeds, which constitute the core mechanisms for GW production \cite{Breitbach2018,ref5,ref8}.

The importance of cosmological phase transitions extends beyond gravitational wave phenomenology; such phenomena may be capable of addressing essential questions in cosmology, such as the nature of dark matter (DM), the origin of the baryon-antibaryon asymmetry, and potential unobserved forces or symmetries in the universe \cite{ref5,ref6,ref14}. FOPTs in dark sectors are particularly interesting since they could occur at energy scales inaccessible to current particle physics experiments, hence providing insights into physics beyond existing collider constraints \cite{ref1,ref5,ref7} -- even though complementary insights might also derive from collider experiments where, for instance, Higgs portal interactions offer observable consequences of dark sectors, linking terrestrial analyses to cosmological phenomena \cite{ref1,ref5,ref24}.

The notion of a ``dark sector'' refers to a range of particles and fields not described by the Standard Model and generally weakly coupled to it, or not coupled at all. These sectors could harbor FOPTs that produce gravitational waves observable at the nanohertz frequency range, a theory supported by multiple models such as those involving additional SU($N$) gauge groups or scalar dynamics \cite{Breitbach2018,ref3,ref5,ref7,ref11}. Some models predict phase transitions in fully secluded hidden sectors, which could carry different temperatures or thermodynamic properties relative to the visible sector, potentially modifying the effective number of relativistic degrees of freedom during their occurrence \cite{Breitbach2018,ref3}.


Despite promising FOPT realizations, other scenarios such as supermassive black hole mergers or cosmic string dynamics also provide viable explanations for the observed nanohertz signals \cite{ref4,ref16,ref17}, or may contribute in addition to the contribution from FOPts. Current data does not unambiguously point to any preferred origin for the nanoHertz signal, however \cite{ref16,ref17}. The field remains an active area of debate and research, driven by the potential to bridge observed cosmic phenomena with theoretical extensions of the Standard Model \cite{ref12,ref17,ref21}.

A prominent class of models within the framework of dark sector FOPT focuses on dark $U(1)$ gauge symmetries, which are minimal extensions of the Standard Model (SM) that naturally admit mechanisms for FOPTs and gravitational wave generation.

Theoretical studies of gravitational waves from FOPTs in dark $U(1)_D$ models have demonstrated their ability to produce magnetic-scale transitions at the low temperatures $T_\star \sim 10~\mathrm{MeV}$, required for PTA-compatible nanohertz signals [6, 7]. These models extend the Standard Model by introducing a dark gauge boson $Z_D$ and a dark Higgs scalar field $\phi_D$, with the dynamics of the first-order phase transition driven by the spontaneous breaking of the $U(1)_D$ symmetry. The scalar potential, typically parameterized with cubic or quartic terms, not only induces the FOPT but also determines the bubble nucleation properties, such as bubble wall velocities $v_w$ and the relative energy partitioning into bubble collisions, sound waves, and turbulence. These factors collectively shape the gravitational wave spectrum \cite{refa6}.

Dark $U(1)$ models offer several attractive features for explaining the PTA signal. The presence of a dark gauge boson $Z_D$ as a mediator between the hidden sector and the Standard Model provides a minimal portal for coupling, such as through kinetic mixing with the photon, allowing limited but phenomenologically relevant interactions with the visible sector. Studies such as Ref.~\cite{refa6} explicitly analyze gravitational wave signals in the context of a thermally decoupled $U(1)_D$ gauge sector, where reheating temperatures $\sim$ MeV  are critical for matching the spectral frequency peak and amplitude $\Omega_{\rm GW}$ to PTA observations. This setup avoids cosmological constraints, such as excessive entropy injection into the visible sector, and prevents disrupting Big Bang Nucleosynthesis (BBN) or Cosmic Microwave Background (CMB) observables.

A related study \cite{refa7} extends the minimal $U(1)_D$ model by incorporating vector dark matter candidates. In this scenario, dark sector gauge bosons acquire mass through the same $U(1)_D$-breaking FOPT, providing both the stochastic gravitational wave background and viable dark matter candidates. This model produces gravitational wave spectra consistent with PTA data while simultaneously explaining the relic abundance of dark matter. Constraining the model to PTA-observable frequency and amplitude ranges requires precise tuning of phase transition parameters, such as the strength of the transition $\alpha$ and the inverse duration $\beta/H$. For example, these studies find that a sufficiently strong phase transition $\alpha > 0.1$ and a relatively slow completion rate $1 \leq \beta/H \leq 100$ are required to consistently produce gravitational wave signals observable at PTA sensitivities \cite{refa6, refa7}.

While dark $U(1)_D$ models achieve compatibility with PTA-observable gravitational wave signals, they also exhibit unique phenomenological testability. The simplest realization of $U(1)_D$ models involves dark Higgs bosons that can mix with the Standard Model Higgs via a Higgs portal. Studies such as \cite{refa4} have explored this mixing as a potential indirect test of first-order phase transitions in the dark sector. For example, precision Higgs measurements at colliders (e.g., the Large Hadron Collider (LHC) or future lepton colliders) could probe invisible decays of the Standard Model Higgs into lighter dark sector particles, thereby constraining the $U(1)_D$-breaking dynamics and the associated gravitational wave signals. Furthermore, such models allow for predictions of associations between gravitational wave characteristics and specific experimental observables, enhancing multi-messenger prospects.

In addition to minimal models, dark $U(1)$ extensions can naturally connect to other cosmological phenomena. For example, modified thermal histories, such as freeze-in or freeze-out mechanisms for the dark Higgs and dark gauge boson, influence the effective degrees of freedom $g_{\rm eff}$ and result in slight shifts to the gravitational wave peak frequency and amplitude \cite{refa3, refa6}. Decoupled dark sectors also evade cosmological bounds on the relativistic energy density $N_{\rm eff} \lesssim 3.5$, provided that the $U(1)_D$ transition reheats only modestly \cite{refa1,refa6, refa8}. More complex cosmological scenarios, such as dark reheating or multiple transitions within the same $U(1)_D$ framework, could further enhance the model's richness and observability \cite{refa3}.

The history of the universe prior to Big Bang Nucleosynthesis (BBN) is only indirectly constrained by cosmological and astrophysical observables, allowing for the possibility of departures from the standard radiation-dominated expansion history during this epoch. Modifications to the pre-BBN Hubble expansion rate, whether through the addition of exotic energy components (e.g., scalar fields, dark radiation, or early decaying particles) or through extensions of general relativity (e.g., scalar-tensor gravity, brane-world cosmologies), profoundly impact the evolution of thermal relic abundances and other cosmological observables. These changes alter the freeze-out and freeze-in processes that, for instance, determine the relic densities of dark matter and other particle species by modifying the interplay between interaction rates and the Hubble expansion rate. 

A range of studies have explored how modified expansion rates affect thermal relic production. For example, faster-than-standard expansions due to kination-like scalar field models $ \rho_\phi \propto a^{-(4+n)} $ with $n > 0$ directly increase the Hubble rate prior to BBN, shortening the thermalization timescale and suppressing freeze-in relic production. Such scenarios also enhance freeze-out abundances by reducing the duration of particle annihilations, a phenomenon extensively analyzed in works introducing ``relentless dark matter'' \cite{refb1, refb7, refb8, refb18}. In contrast, slower-than-standard expansion rates often arise in early matter-dominated (EMD) cosmologies dominated by long-lived massive particles, which delay freeze-out processes and allow for entropy injection upon particle decay. These effects have been systematically studied in the context of pre-BBN reheating scenarios \cite{refb4, refb13, refb23}. Scalar-tensor gravity models and other modifications to the Friedmann equation provide additional degrees of freedom for modeling these early expansion histories. These frameworks have demonstrated the ability to significantly alter thermal relic evolution, such as reannihilation phenomena or relic abundance enhancement under certain parameter regimes \cite{refb2, refb3, refb5, refb16, refb19}.

A critical aspect of these models is their compatibility with observational constraints. All modifications to the pre-BBN expansion rate must transition to the standard cosmological evolution by $T \sim 10 \, \text{MeV} $, ensuring consistency with BBN measurements of light element abundances and the effective number of relativistic species $N_{\text{eff}}$ as constrained by the Cosmic Microwave Background (CMB) \cite{refb6, refb9, refb15}. Quantitative studies incorporating these constraints include numerical solutions of the Boltzmann equation coupled to modified Friedmann equations, which have been widely used to derive accurate predictions for thermal relic densities across a variety of non-standard cosmological scenarios \cite{refb1, refb3, refb4, refb7, refb10, refb18}. Furthermore, observational signatures of these modified expansion histories, such as altered annihilation rates and their potential consequences for dark matter indirect detection signals (e.g., gamma rays or neutrinos), provide complementary avenues for testing these hypotheses \cite{refb14, refb23}.

Below, we point out that the dynamics of a strongly FOPT and of the resulting gravitational wave spectrum is tightly coupled to the background cosmology, in ways that we will make clear in what follows. Interestingly, for a FOPT occurring in a background cosmology with a {\it larger} Hubble rate, the resulting gravitational wave spectrum is greatly enhanced. As a result, FOPT that would produce excessively weak gravitational wave signals may actually fit data in the presence of a modified early universe cosmology.

Here, we first study the conditions and implications of explaining the NANOGrav signal with a dark-sector FOPT based on the spontaneous symmetry breaking of a dark, Abelian gauge symmetry; we then further consider a combination of a FOPT and SMBHB mergers. To this end, we utilize the most up-to-date NANOGrav data, as well as best-practice Bayesian analysis tools.

In a simple model for the FOPT we produce an analytical form for the relevant quantities needed to fit the NANOGrav signal, namely the normalization and pivot frequency of the stochastic background produced at the FOPT; we then conduct a posterior analysis recovering a preferred range for the effective model parameters, which we then map back onto the theory parameter space via a state-of-the-art numerical code, {\tt TransitionListener}; We find, with accurate statistical and numerical methods that the reconstructed likelihood contours  indicate that a FOPT would generally need to be harbored in a dark sector featuring either a very large number of dark degrees of freedom compared to the universe's visible sector, or taking place in a modified cosmology with a faster-than-radiation Hubble expansion rate. For the first time, we point out that a FOPT taking place in a modified cosmological setting can result in a very large enhancement to the amplitude of the gravitational wave spectrum, making models otherwise insufficient to fit the PTA/NANOGrav signal viable.



The remainder of the paper is structured as follows: in sec.~\ref{sec:GWs} we describe the generation of GWs in a dark sector FOPT; 
sec.~\ref{sec:param} details the model under consideration and the relevant input parameters, and it explains, as well, the analytical functional form with which we approximate the FOPT stochastic background of GWs which we use to fit the NANOGrav data; sec.~\ref{sec:modifiedcosmo} introduces modified expansion histories at early times, and explains how they affect the FOPT and the resulting gravitational wave yield; sec~\ref{sec:NG} presents our methods to fitting the NANOGrav data, and sec.~\ref{sec:results} details our findings. The concluding sec.~\ref{sec:conclusions} presents our discussion and conclusions, and it outlines future directions.




\section{Gravitational Waves from Cosmological First-Order Phase Transitions}\label{sec:GWs}

GW signals from FOPTs in a dark sector are driven by key parameters such as the latent heat ($\alpha$), transition duration ($\beta^{-1}$), and the thermal state of the dark sector, characterized by its temperature ($T_{\mathrm{DS}}$) \citep{Feng2024, Breitbach2018, Ertas2021, Fairbairn2019}. The dominant mechanisms for GW production include bubble collisions, sound waves in the plasma, and magnetohydrodynamic (MHD) turbulence, with the relative contributions determined by the dynamics of bubble walls and the degree of energy transfer into different channels \citep{Wang2022, Croon2018}.

The latent heat ($\alpha$) plays a critical role in determining the overall strength of the GW signal, while the normalized inverse transition rate ($\beta/H$) controls both the peak frequency and amplitude of the signal. A stronger phase transition (larger $\alpha$) tends to amplify the GW signal, while smaller values of $\beta/H$ (longer-duration transitions) shift the spectrum to lower frequencies, which may be more detectable for ground- or space-based GW observatories such as LISA or DECIGO \citep{Feng2024, Breitbach2018, Wang2022}.

The dark sector temperature, $T_{\mathrm{DS}}$, has a significant impact on both the dynamics and the detectability of phase transitions in hidden sectors. Studies indicate that dark sectors decoupled from the visible sector can evolve independently, with higher $T_{\mathrm{DS}}$ amplifying the GW amplitude but also increasing the effects of signal dilution via entropy transfer \citep{Ertas2021, Fairbairn2019}. Conversely, low-temperature ($T_{\mathrm{DS}} \sim \text{MeV}$) transitions result in spectral features that may lie in unique frequency ranges, including nano-Hertz signals we are focusing on here \citep{Bringmann2023, Fairbairn2019}.

We note that the thermal evolution of dark sectors, including the role of $T_{\mathrm{DS}}$ and entropy transfer, may provide a unique opportunity to distinguish dark sector transitions from visible sector analogues, highlighting the importance of understanding the interplay between thermodynamics and particle interactions in such scenarios \citep{Croon2018, Ertas2021}.

\section{A Dark Photon - Dark Higgs Model}\label{sec:param}
The toy model we adopt here follows what considered in detail in Ref.~\cite{Ertas2021}, to which we refer the Reader for additional details. The Standard Model (SM) is augmented by a new Abelian gauge group $U(1)_D$ under which only a complex, ``Dark'' Higgs field $\Phi=(\phi+i\varphi)/\sqrt{2}$ is charged, with all other SM fields neutral. The Lagrangian of the SM is then augmented by a dark-sector Lagrangian
\begin{equation}
{\cal L}\in |D_\mu \Phi|^2+|D_\mu H|^2-\frac{1}{4}A^\prime_{\mu\nu}A^{\prime\mu\nu}-\frac{\varepsilon}{2}A^\prime_{\mu\nu}B^{\mu\nu}-V(\Phi,H),
\end{equation}
where we have included the SM Higgs $H$,  where the covariant derivative
\begin{equation}
D_\mu\Phi=\left(\partial_\mu+ig_D A^\prime_\mu\right)\Phi,
\end{equation}
and, finally, where the field-strength tensor $X_{\mu\nu}\equiv\partial_\mu X_\nu-\partial_\nu X_\mu$ ($A^\prime$ is the dark photon and $B$ the hypercharge gauge boson).

The tree-level scalar potential reads
\begin{equation}
V_{\rm tree}(\Phi,H)=\mu^2\Phi^*\Phi+\lambda(\Phi^*\Phi)^2-\mu_H^2H^\dagger H+\lambda_H(H^\dagger H)^2+\lambda_p(\Phi^*\Phi)(H^\dagger H).
\end{equation}
Assuming $\varepsilon$ and $\lambda_p$ to be small enough for the dark sector not to thermalize with the SM, effectively the model is defined in terms of the couplings $\lambda,\ g$ and the mass-scale $\mu$. The mass terms read \cite{ref10ofTL}
\begin{eqnarray}
m^2_{A^\prime}&=&g^2\phi^2,\\
m_\phi^2&=&-\mu^2+3\lambda\phi^2,\\
m_\varphi^2(h,\phi)&=&-\mu^2+\lambda \phi^2.
\end{eqnarray}
After symmetry breaking, the minimum for the potential of $\phi$ lies at $v=\mu/\sqrt{\lambda}$, resulting in the dark photon mass $m_{A^\prime}=g\ v$ and $m_\phi=\sqrt{2\lambda} v$. While the lifetime of the dark Higgs is, in a minimal setup, set by $\lambda_p$, {\tt TransitionListener} takes a model-agnostic stand and adopts a generic lifetime $\tau$. Also, in the model we  assume a dark sector to visible sector temperature ratio $\xi$.

In sum, the model under consideration is controlled phenomenologically by the dark Higgs quartic coupling $\lambda$, the gauge coupling $g$, the vev $v$, the dark Higgs lifetime $\tau$, and the temperature ratio $\xi$.

The model under consideration is found below to produce a gravitational wave amplitude insufficient to explain the NANOGrav data. We therefore consider two possibilities: the phase transition occurs in a modified expansion phase of the early universe (see the next section), or the dark sector consists of numerous {\em identical} copies of the model described here, and the gravitational wave yield scales linearly with this number of dark sector copies, which we indicate with $N_{\rm DS}$. The latter case may appear in contexts were dark sectors include a large number of degrees of freedom \cite{Ewasiuk:2024ctc}, such as ``clockwork''  \cite{Giudice:2016yja}, dynamical dark matter \cite{Dienes:2011ja,Dienes:2011sa}, and ``stasis'' \cite{Dienes:2021woi} scenarios.

\section{Gravitational wave signal with a modified Hubble rate}\label{sec:modifiedcosmo}

The universe's thermal history is strongly constrained by observations at late times. However, not much is known at times prior to the synthesis of light elements, occurring at around $t\sim $ few seconds and temperatures in the few MeV. Specifically, the expansion rate of the universe at higher temperatures/earlier times may have well departed from the straightforward extrapolation of radiation domination back to the epoch of reheating. Studies have contemplated both a {\it slower} expansion rate, for instance if the universe is dominated by a matter species that decays at late times, and {\it faster}, such as if the universe is dominated at early times by a fast-rolling scalar field, i.e. a field whose energy density is dominated by its kinetic energy term, $\dot \phi^2\gg V(\phi)$, a possibility sometimes dubbed ``kination''. We give several additional examples in the Introduction. Here, we discuss how a modified expansion history affects a FOPT and the resulting gravitational wave spectrum.

If the energy density of the species that dominates in the early universe (which refers here to the period after reheating and before BBN) has an energy density $\rho\sim T^{4+n}$, the the Hubble rate
\begin{equation}
    H=\frac{\sqrt{\rho}}{\sqrt{3}M_{\rm Pl}}\sim T^{2+n/2}.
\end{equation}
Indicating with $T_r$ (following the notation of \cite{refb1}) the temperature at which the energy density of the additional component equals that of radiation, and neglecting the change in the number of degrees of freedom at early times before and after $T_r$, we can write
\begin{equation}
    H\sim H_{\rm rad}\left(\frac{T}{T_r}\right)^{n/2}\quad T\gg T_r,
\end{equation}
where $H_{\rm rad} $ is the Hubble rate in a radiation dominated universe.

The normalization of the gravitational wave background from a strongly first-order phase transition at production is proportional to the factor $\Omega_{\rm GW}\sim (H/\beta)^q$, with $q\simeq1$, with $\beta$ the inverse timescale of the phase transition duration. Additionally, the amplitude of the gravitational wave signal today compared to that at production is
\begin{equation}
    \Omega^0_{\rm GW}(f)=\Omega_{\rm GW}\left(\frac{a}{a_0}\right)^4\left(\frac{H}{H_0}\right)^2,
\end{equation}
where ``0'' indicates the present time. Compared to the radiation domination case,
\begin{equation}
   \left(\frac{a}{a_0}\right)_{\rm rad}\Rightarrow \frac{T_r}{T}\left(\frac{a}{a_0}\right)_{\rm rad},
\end{equation}
where $T$ indicates the phase transition temperature; similarly,
\begin{equation}
   \left(\frac{H}{H_0}\right)_{\rm rad}\Rightarrow \left(\frac{T}{T_r}\right)^{2+n/2}\left(\frac{H}{H_0}\right)_{\rm rad}.
\end{equation}
Finally, compared to the standard case, the phase transition normalization changes as
\begin{equation}
\Omega^0_{\rm rad}\Rightarrow     \Omega^0_{\rm rad}\left(\frac{a_0}{a}f\right) \left(\frac{T}{T_r}\right)^{(n/2)}\left(\frac{T_r}{T}\right)^4\left(\frac{T}{T_r}\right)^{4+n}=\left(\frac{T}{T_r}\right)^{3n/2}.
\end{equation}
For a phase transition occurring at a temperature $T\sim 10$ MeV, and for, for instance, a kination domination scenario, where $n=2$, with $T_r\sim 1$ MeV, the enhancement can be of up to $(T/T_r)^3\sim 1000$. Note that however the shift in frequency 
\begin{equation}
f_{\rm rad}\Rightarrow \frac{a_0}{a}f_{\rm rad}    
\end{equation}
will generate a different preferred vev as well.


\section{Numerical Methods and Applying Constraints from the NANOGrav Collaboration} \label{sec:NG}

The NANOGrav collaboration, along with several other pulsar timing array (PTA) collaborations, have recently presented evidence for the existence of a common red noise process in their pulsar data which obeys the Hellings and Down (HD) correlation pattern \cite{Hellings1983}. The HD correlation pattern is considered  to be the telltale sign of a stochastic background of gravitational waves in the PTA datasets, leading many to interpret this signal as a GWB (see e.g. \cite{ref1,ref6,ref18,ref4,ref16,ref17}). One of the simplest explanations for the origin of this GWB is the superposition of nHz-frequency GWs from a population of SMBHBs, which is expected to generate a power-law spectrum of the form \cite{ref4}:

\begin{equation} 
h^2 \Omega_{\rm GW}(f) = \frac{2 \pi^2}{3 H_{100}^2} f^2 h_c^2(f) = \frac{2 \pi^2}{3 H_{100}^2} A^2 f_{\rm yr}^2 \left( \frac{f}{f_{\rm yr}} \right)^{5-\gamma}
\label{power-law}
\end{equation}\\ where $f_{yr} = 1{\rm yr}^{-1}$,$ H_{100} = H/100$ km/s/Mpc, and $\gamma = \frac{13}{3}$ in the case of SMBHBs evolving purely via GW emission. However, numerous alternative explanations for the origin of the GWB have been suggested as well \cite{ref6,ref18,ref4,ref16,ref17,ref18}. Broadly speaking these alternative sourcing mechanisms involve cosmological, rather than astrophysical, processes, and many of them are able to provide better fits to the GWB power-spectrum than the astrophysical case. One such example of a cosmological signal is a GWB sourced by a DSPT in the early universe. This scenario has been investigated before e.g. in \cite{Bringmann2023}. In our work, we explore two potential modifications to the GW production from DSPTs and demonstrate that a potentially weak DSPT, such as in the model under consideration, can still lead to a large-enough amplitude  to explain the NANOGrav signal.

Owing to the computational complexity of generating precise GW-spectra from a given set of physical phase transition parameters, we first identified a phenomenological model which retains the qualitative structure of most maximal DSPT GW spectra while being generated from a much simpler set of parameters, namely the spectrum's peak amplitude $\Omega_{\star}$ and the position of this peak $f_{\star}$. Following \cite{Romero-Rodriguez:2022ebc}, the form of this phenomenological DSPT model is as follows:

\begin{equation} 
h^2 \Omega_{\rm GW}(f) = \Omega_{\star}\left(\frac{f}{f_{\star}}\right)^{n_{1}}\left(1+\left(\frac{f}{f_{\star}}\right)^{\Delta}\right)^{(n2-n1)/\Delta},
\label{DSPT-Phenom}
\end{equation}
where $n_{1}=3$, $n_{2}=-1$, and $\Delta= 4$ for the particular case we are interested in. We use the PTA data analysis suite {\tt PTArcade} \cite{Mitridate:2023oar} to perform our fits and obtain posterior distributions over our model parameters. {\tt PTArcade} allows for an efficient computation of ideal model parameters by performing Bayesian Monte Carlo Markov Chain (MCMC) searches over the model parameter space, either comparing the models directly to pulsar timing data or to the correlated timing noise residual power spectrum. In our case, we used {\tt PTArcade} to fit our model against the NG15yr HD-correlated timing noise residual power spectrum \cite{2023PhRvD.108j3019L}.

We employed {\tt TransitionListener} to generate a library of GWB spectra which we used for translating between our phenomenological constraints on $f_{\star}$ and $\Omega_{\star}$ and the physical parameters of interest to us. {\tt TransitionListener} is based on {\tt CosmoTransitions} \cite{Wainwright2012CosmoTransitions} which was designed to analyze finite-temperature cosmological phase transitions driven by single or multiple scalar fields. It can trace the global and local minima of a given effective potential, identify possible phase transitions between  minima as a function of temperature, and calculate bounce actions and bubble profiles. This allows for a flexible calculation of the spectral shape of GWBs produced in FOPTs given different physical input parameters.

{\tt TransitionListener} takes five input parameters by default, and we manipulated two parameters to generate our template spectra: the ratio of photon bath temperature to the dark sector temperature at the time of nucleation of the DSPT $\xi$ and the vacuum expectation value of the dark sector, $v$. The other parameters $\lambda, g,$ and $\gamma$ were fixed at default values of $0.2$ GeV, $1$, and $10^{-16}$ GeV respectively, as these parameters produced GW spectra which exhibited the maximal amplitude peaks, while also retaining a smooth spectral shape that could be easily approximated with a phenomenological model (discussed above). In our study, we generated a library of 50 sample GW spectra which were log-uniform in $v$ space from $10^{-6}$ GeV to $10^{10}$ GeV.

Once we had our library of GWB spectra from {\tt TransitionListener}, we scanned over our model parameters (such as $\xi$, $v$, $T_{r}$, or $N_{DS}$), noting the values of $f_{\star}$ and $\Omega_{\star}$ each set of parameters produced, effectivley creating a map between the physical parameter space and the phenomenological one for which we obtained constraints. To find the goodness of fit for each set of parameters based on their resulting spectral peak, we calculated the Mahalanobis distance of this peak from the center of the $(f_{\star},\Omega_{\star})$ MCMC distribution. This method of translating from a generic $f_{\star}$ and $\Omega_{\star}$ to a specific number of standard deviations from the center of the MCMC assumes the MCMC contours are a multivariate Gaussian, which we found to hold true to good approximation within the first 2-3 standard deviations.

\begin{figure}[!t]
    \centering
     \mbox{\includegraphics[height=6.0cm]{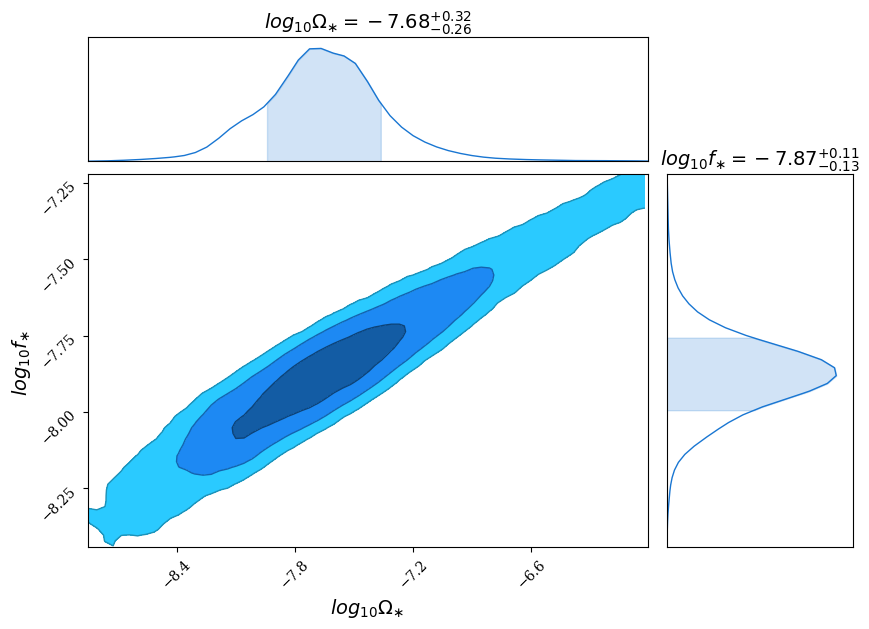}\qquad \includegraphics[height=6.0cm]{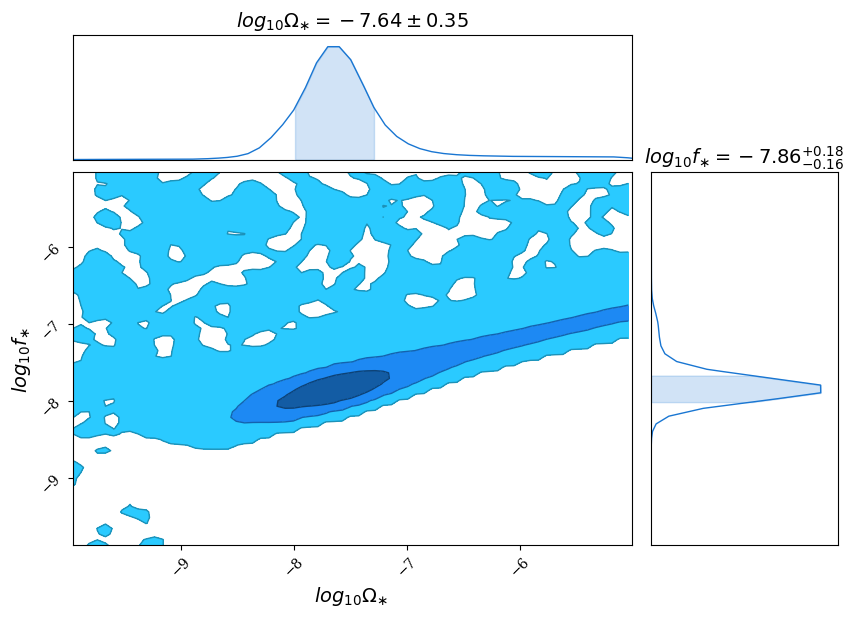}
    }\\
    \caption{MCMC posteriors for our phenomenological DSPT model (left) and DSPT+SMBHB model (right) when fit to the NG15yr HD posterior spectrum. Contours show the $1\sigma$, $2\sigma$, and $3\sigma$ constraints.}
    \label{fig:fstar_omegastar_2sigma}
\end{figure}


\section{Results} \label{sec:results}

\begin{figure}[!t]
    \centering
     \mbox{\includegraphics[height=8.0cm]{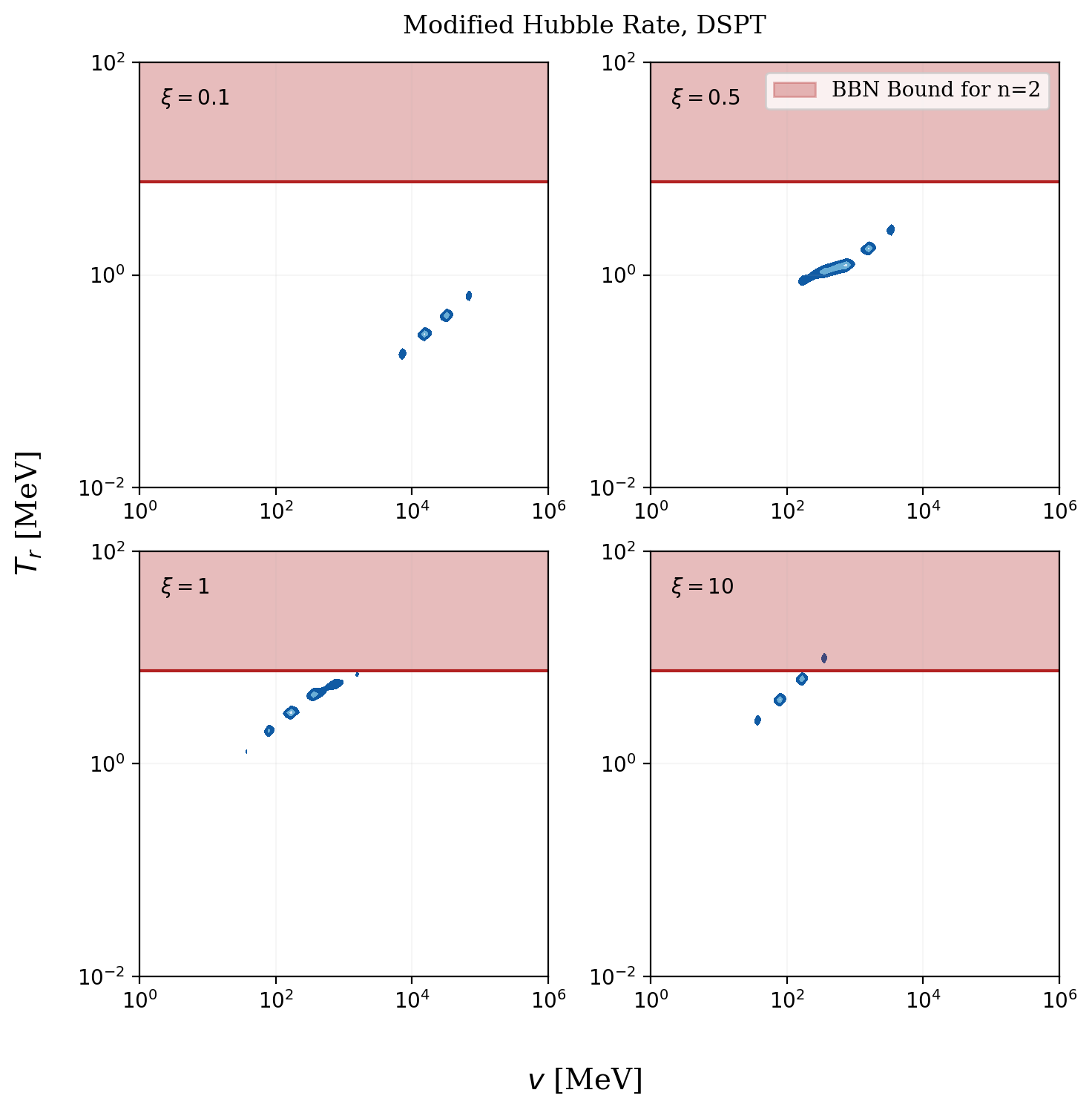}}\qquad
    \includegraphics[height=8.0cm]{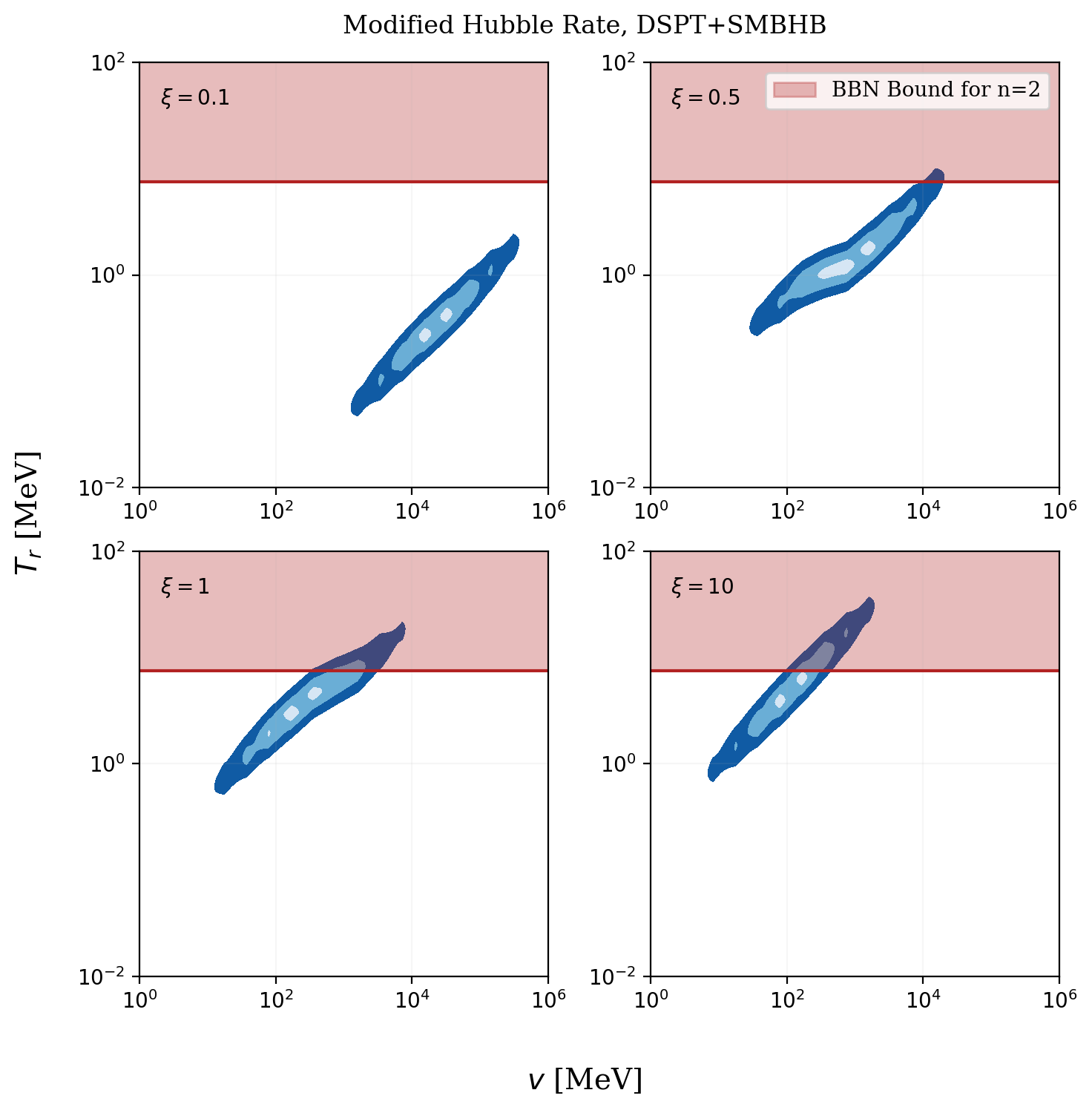}
    \\
    \caption{Posteriors for $v$ versus $T_{r}$ translated from the MCMC posterior constraints on $f_{\star}$ and $\Omega_{\star}$ given for $\xi= 0.1, 0.5, 1,$ and $10$. Left: constraints given a DSPT-only phenomenological model. Right: constraints given a model combining signals from a DSPT and a power-law GWB sourced by SMBHBs.}
    \label{fig:vev_Tr}
\end{figure}

\begin{figure}[!t]
    \centering
     \mbox{\includegraphics[height=7.5cm]{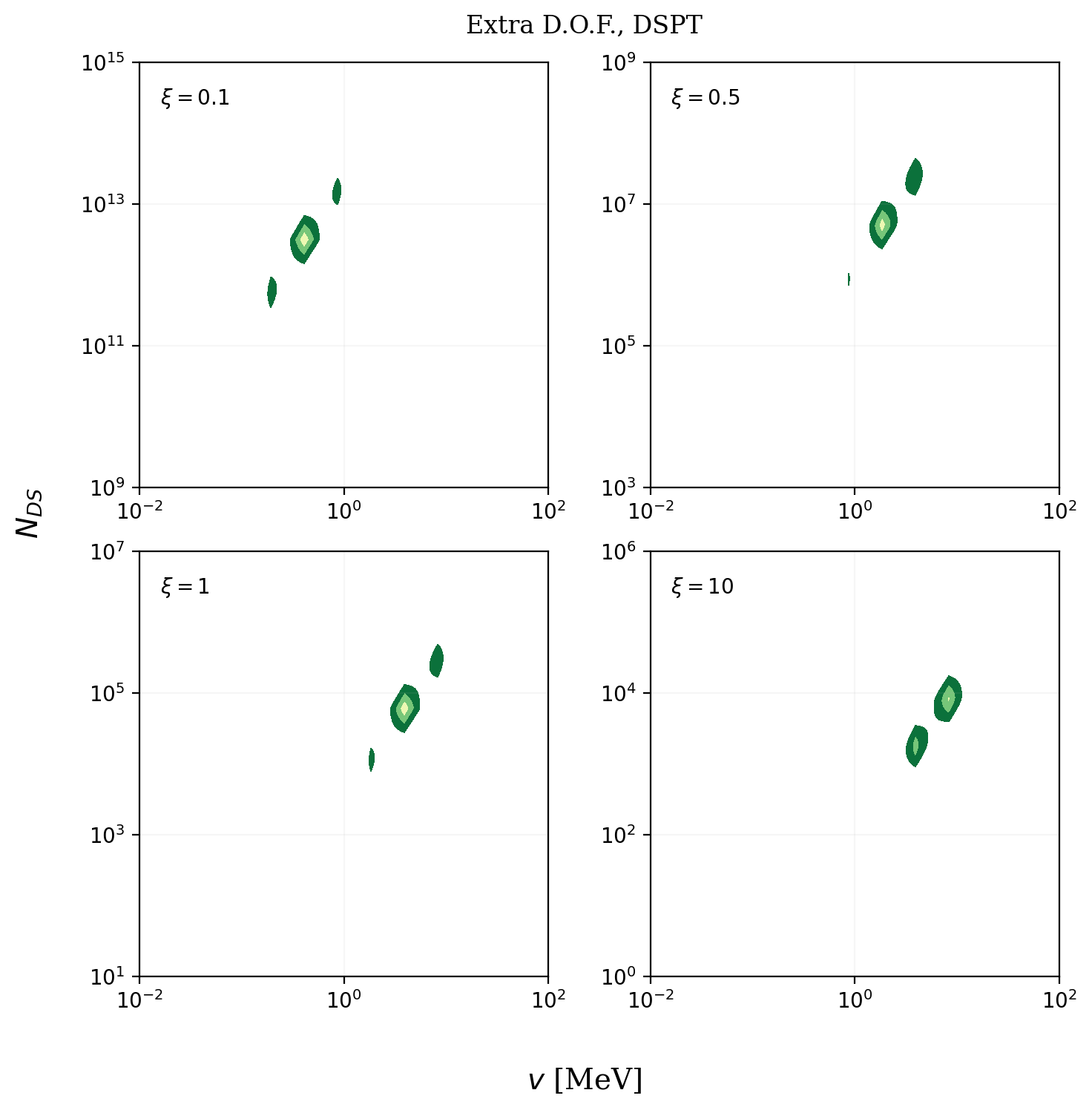}}\qquad \includegraphics[height=7.5cm]{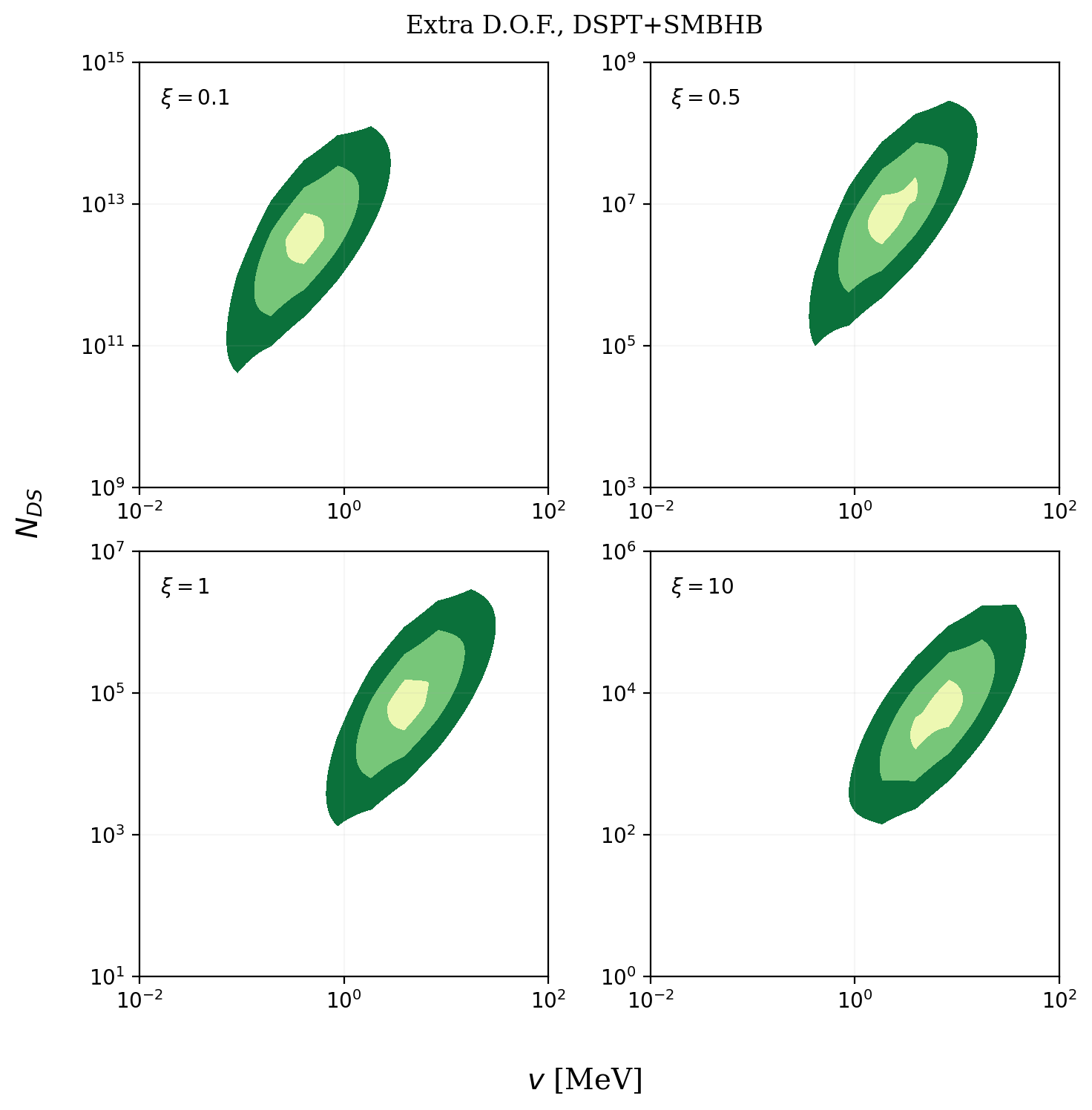}
    \\
    \caption{Same as Fig. ~\ref{fig:vev_Tr} but for parameters $v$ and $N_{\rm DS}$ in a model where the DSPT is boosted by extra degrees of freedom in the dark sector.}
    \label{fig:vev_ndf}
\end{figure}

The constraints on the phenomenological DSPT model when fit to the NANOGrav 15-year dataset are shown in Fig.~\ref{fig:fstar_omegastar_2sigma}. These constraints on $(f_{\star},\Omega_{\star})$ are translated to preferred values for $(T_{r},v)$ in Fig.~\ref{fig:vev_Tr} via the process described in ~\ref{sec:NG}. Because $\xi$ is unlikely to surpass $\mathcal{O}(10)$ in physically motivated scenarios \cite{Breitbach2018}, we chose to show constraints for $\xi = 0.1, 0.5, 1,$ and $10$. The results from one additional case we considered, in which the DSPT signal is boosted by the presence of numerous degrees of freedom in the dark sector, is shown in Fig.~\ref{fig:vev_ndf}.

For the case of an enhancement to the DSPT-only signal by a modified Hubble rate in the early universe, our constraints on $f_{\star}$ and $\Omega_{\star}$ lead to a preference for a reheating temperature between around 0.1 and 10 MeV, and a $v$ between 10 MeV and 100 GeV, depending on the choice of $\xi$. In the particular model we chose, we set $n=2$, leading to a constraint from Big Bang Nucleosynthesis (BBN) of $T_{r}\leq 7.5$ MeV, which is shown in red (see \cite{DEramo:2017gpl} for more details on this constraint). Most of the preferred region of parameter space is safe from this constraint except for at the high end of $\xi$ and $v$ values. 

The $3\sigma$ contours for the cases of $\xi=0.1, 0.5,$ and $1$ are shown in the left panel of Fig.~\ref{fig:vev_Tr}. The addition of a GWB signal from SMBHBs enables a larger range of parameters to fit well to the signal, although the $3\sigma$ contours of the phenomenological model show that there is still a preference for a strong contribution from the additional DSPT signal. This explains why our $3\sigma$ contours for $T_r$ and $v$ remain somewhat tightly constrained around non-zero values even in the DSPT+SMBHB case, as depicted in the right panel of Fig.~\ref{fig:vev_Tr}. Of course, it is also possible for the SMBHB signal to fully dominate, which allows the DSPT signal to take on nearly arbitrary $f_{\star}$ values so long as $\Omega_{\star}$ remains sufficiently low. This flexibility generally permits lower $T_{r}$ and $v$ values. It is also possible that the species dominating in the early universe obeys an equation of state with $w > 1$. Because the energy density of this species scales like $\rho\sim T^{4+n}$ where $n=3w-1$, this would lead to the energy density scaling more strongly with T. However, we found that increasing $n$ generally pushed the preferred values of $T_{r}$ closer to (or beyond) constraints from BBN and decreased the size of viable parameter space.

For the case of an enhancement to the DSPT signal by the existence of numerous dark sectors, $N_{\rm DS}$, our constraints on $f_{\star}$ and $\Omega_{\star}$ lead to a preference for an enormous, physically unrealistic number of dark sectors -- above $10^{12}$ for the case of $\xi=0.1$ and still above $10^{2}$ for $\xi=10$. Across these three scenarios for $\xi$, the preferred $v$ range remains in a small window between $0.3$ and $3$ MeV. The specific $3\sigma$ constraints are shown in the left panel of Fig.~\ref{fig:vev_ndf}. As before, the DSPT+SMBHB case slightly extends the range of the $3\sigma$ contours. Of course, in the limiting case where the DSPT contribution to the nHz GWB goes towards zero, the parameters $N_{\rm DS}$ and $v$ could extend arbitrarily close to zero.

\section{Discussion and Conclusions}\label{sec:conclusions}

Here we have examined the potential for a dark sector phase transition (DSPT) which may be insufficiently ``loud'' to explain the spectrum of the gravitational wave background (GWB) measured in the recent NANOGrav 15-year dataset, to be boosted in the context of two scenarios: a modified expansion history in the early universe, or in the presence of a large number of copies of a dark sector. We used a phenomenological model describing the maximal GW signal produced by such a phase transition consisting of only two parameters, $f_{\star}$ and $\Omega_{\star}$. We then fit to physical parameters $T_{r}, v, $ and $\xi$ and found that the modified Hubble rate enhancement of a DSPT can produce signals that fit the NANOGrav signal, despite this {\em not} being the case in a standard cosmological story. For this to be the case, we find we would need to have a low reheating temperature scale around $10^{-1}$ to $10^{1}$ MeV and DSPT vacuum expectation values $v$ between about $10^{1.5}$ and $10^{5.5}$ MeV, depending on the choice of $\xi$, the dark-to-visible sector temperature ratio. For the modified Hubble rate scenario, we also found that for lower $\xi$ values the preferred $v$ values increase towards larger energies, while the most likely reheating temperature scale values $T_{r}$ shrink lower. This is because for lower values of $\xi$, the $v$ values which are able to produce a signal with a peak in the right frequency range shift lower, typically corresponding to a signal with a lower overall amplitude. This increases the amount of signal boosting required, leading to lower preferred values of $T_{r}$. We also considered different equation of state $w$ for the species that dominates the energy density of the universe at early times, and found that the corresponding effects are not as dramatic as for $w=1$. 

For an alternative scenario for boosting the DSPT signal such as a large number of dark sectors, we also find we can boost the GW signal to match the NANOGrav signal, but find we would need to have $N_{\rm DS} \geq 10^{3}$ while obtaining $v \leq 10^{1.5}$ MeV.


In both cases, we find that pairing the DSPT GW signal with an astrophysical GWB from a population of supermassive black hole binaries unsurprisingly gives us greater flexibility, and a broader parameter space for the values of $N_{\rm DS}, T_{r},$ and $v$ that would produce a result compatible with the NANOGrav data.

\section*{Acknowledgements} \label{sec:acknowledgements}
   This material is based upon work supported in part by the U.S. Department of Energy grant number de-sc0010107 (SP).

\bibliographystyle{unsrt} 
\bibliography{bib} 

\end{document}

%% file: preamble.tex
\usepackage{amsthm}
\usepackage{mathtools}
\usepackage{physics}
\usepackage{xcolor}
\usepackage{graphicx}
\usepackage[left=23mm,right=13mm,top=35mm,columnsep=15pt]{geometry} 
\usepackage{adjustbox}
\usepackage{placeins}
\usepackage[T1]{fontenc}
\usepackage{lipsum}
\usepackage{csquotes}